\documentclass[aps,prl,10pt,twocolumn,amsmath,amssymb,superscriptaddress]{revtex4-2}
\usepackage[]{graphicx}
\usepackage{comment}
\newcommand{\del}{\partial}

\usepackage{color}
\usepackage{hyperref}
\hypersetup{breaklinks=true}

\newcommand{\tr}{\text{tr}}

\newcommand {\nn} {\nonumber}
\newcommand{\bbR}{{\mathbb R}}

\begin{document}
\begin{titlepage}

  \title{Defining the type IIB matrix model without breaking
    Lorentz symmetry}

\author{Yuhma A{\sc sano}}
\email{asano@het.ph.tsukuba.ac.jp}
\affiliation{Institute of Pure and Applied Sciences, University of Tsukuba,\\
1-1-1 Tennodai, Tsukuba, Ibaraki 305-8571, Japan}
\affiliation{Tomonaga Center for the History of the Universe,
  University of Tsukuba,\\
1-1-1 Tennodai, Tsukuba, Ibaraki 305-8571, Japan}
  
\author{Jun N{\sc ishimura}}
 \email{jnishi@post.kek.jp}

 \author{Worapat P{\sc iensuk}}
\email{piensukw@post.kek.jp}

\author{Naoyuki Y{\sc amamori}}
\email{yamamori@post.kek.jp}

\affiliation{KEK Theory Center,
Institute of Particle and Nuclear Studies,\\
High Energy Accelerator Research Organization,\\
1-1 Oho, Tsukuba, Ibaraki 305-0801, Japan}
\affiliation{Graduate Institute for Advanced Studies, SOKENDAI,\\
1-1 Oho, Tsukuba, Ibaraki 305-0801, Japan}


\date{April 22, 2024; preprint: UTHEP-787, KEK-TH-2617}

\begin{abstract}
  The type IIB matrix model is a promising nonperturbative formulation
  of superstring theory, which may elucidate the emergence of (3+1)-dimensional
  space-time. However, the partition function is divergent
  due to the Lorentz symmetry, which is represented by a noncompact group.
  This divergence has been regularized conventionally
  by
  %
  introducing some infrared cutoff,
  which breaks the Lorentz symmetry.
  Here we point out, in a simple model, that Lorentz invariant observables
  become classical as one removes the infrared cutoff and that
  this ``classicalization'' is actually an artifact of the Lorentz symmetry
  breaking cutoff. In order to overcome this problem, we propose a natural
  way to ``gauge-fix'' the Lorentz symmetry in a fully nonperturbative manner.
  Thus we arrive at a new definition of the type IIB matrix model,
  which also enables us to perform numerical simulations in such a way
  that the time-evolution can be extracted from the generated configurations.
\end{abstract}
\maketitle
\end{titlepage}
\textit{Introduction.---}
It is widely believed that superstring theory is
the fundamental theory that describes our Universe including quantum gravity.
The type IIB matrix model \cite{9612115}
(or the Ishibashi-Kawai-Kitazawa-Tsuchiya model)
is a promising candidate of a nonperturbative
formulation of superstring theory, which may play a crucial role
analogous to the lattice gauge theory in understanding its nonperturbative dynamics.
In particular,
it is possible
that (3+1)-dimensional space-time emerges
from (9+1)-dimensional space-time, in which superstring theory is formulated.
While the action of the model was given in the original paper,
the partition function actually diverges
due to the Lorentz symmetry,
which is represented by a noncompact group.
This divergence has been dealt with conventionally
by introducing some infrared cutoff, which breaks Lorentz symmetry.
See Refs.~\cite{Anagnostopoulos:2022dak,Brahma:2022ikl,Klinkhamer:2022frp,Steinacker:2024unq}
for related reviews and a textbook.

In this Letter,
we first point out, in a simple model, that
Lorentz invariant observables become classical as one removes the cutoff.
This ``classicalization'' is actually an artifact of the Lorentz symmetry
breaking cutoff and
it can be understood by considering the Hessian around the saddle point.
Since the Hessian transforms covariantly under the Lorentz transformation,
the fluctuations around the saddle point is boosted
for the boosted saddle point.
However, in the presence of the cutoff,
the fluctuations are effectively restricted to the directions tangential to the
cutoff surface.
This eliminates
quantum fluctuations
and
causes the classicalization.

Motivated by this new insight,
we propose to make the
Lorentz symmetric model well-defined
by ``gauge-fixing'' the Lorentz symmetry in a fully nonperturbative manner. 
Unlike the model
with the cutoff,
Lorentz invariant
observables do not classicalize, which clearly
confirms that
the classicalization is indeed an artifact of the Lorentz symmetry breaking cutoff.  
Generalizing
this idea,
we propose a new definition of the
type IIB matrix model,
which does not suffer from
such artifacts of the Lorentz symmetry breaking cutoff.


\textit{Examples with one Lorentz vector.---}
Before we discuss the type IIB matrix model, which consists of
$(N^2-1)$ Lorentz vectors, where $N$ is the size of the matrices,
it is useful to discuss Lorentz symmetric models with one Lorentz vector.
While the discussion here is quite elementary, it tells us all
the essence of the issues we may encounter in the type IIB matrix model.

First let us consider
the partition function
\begin{align}
  Z  &=
  \int \, dx  \, e^{-S(x)} \ ,
  \quad 
  \label{gaussian-quartic}
  S(x) =  \frac{1}{2} \gamma ( \eta_{\mu\nu}  x_\mu x_\nu + 1 )^2  \ ,
\end{align}
where $\gamma>0$ and $x_\mu \in \bbR$ $(\mu = 0 , 1 , \cdots , d)$.
The Lorentz metric is defined by $\eta_{\mu\nu}= {\rm diag} (-1 , 1 , \cdots , 1)$
and the model \eqref{gaussian-quartic} has Lorentz symmetry
$x_\mu \mapsto {\cal O}_{\mu\nu} x_\nu $, where ${\cal O} \in {\rm SO}(d,1)$.
Repeated indices are summed over.

For simplicity, let us
focus on the large $\gamma$ region,
where the saddle-point analysis is expected to be valid.
There are two types of saddles.
One is
(i) $\eta_{\mu\nu} x_\mu x_\nu = - 1 $
and the other is (ii) $x_\mu=0$.
Saddle points of the first type
are related to each other by Lorentz transformation,
and each of them
contributes equally,
which makes the partition function divergent.
However, this divergence is simply due to the noncompactness
of the Lorentz group.

\textit{Classicalization in the cutoff model.---}
Let us see what happens if one regularizes this model \eqref{gaussian-quartic}
by introducing the Lorentz symmetry breaking cutoff as
\begin{align}
    Z_{\epsilon} &=
    \int \, dx \, e^{ -S(x)
    -  \epsilon (x_0)^2 - \epsilon (x_i)^2 } \ ,
  \label{sym-br-cutoff}
  \end{align}
where $\epsilon$ is the cutoff parameter that is sent to zero later.
Since the action involves quartic terms in $x$, we introduce an auxiliary variable
$k$ and rewrite it into a quadratic form in $x$ as
\begin{align}
  Z_{\epsilon} &=
    \frac{1}{\sqrt{2 \pi \gamma}}
    \int \,  dk
    \, dx \,
  e^{-\frac{1}{2\gamma} k^2 + i k (\eta_{\mu\nu}x_\mu x_\nu +1)
    - \epsilon (x_0)^2 - \epsilon (x_i)^2 } 
    \ .
\label{k-x-integral}
  \end{align}
Note that one can retrieve \eqref{sym-br-cutoff} by integrating out $k$.
Let us then integrate out $x$ in \eqref{k-x-integral}, which yields
\begin{align}
  Z_{\epsilon}
  &=
      \frac{1}{\sqrt{2 \pi \gamma}}
      \int \, dk \, 
  e^{-\frac{1}{2\gamma} k^2 +ik}
   \sqrt{\frac{\pi}{ik + \epsilon}}
   \left( \sqrt{\frac{\pi}{-ik + \epsilon}} \right)^d \nn \\
 &= {\cal N} \int \, dk \, e^{- S_{\rm  eff}(k)}    \ ,
\label{kxx-integrate-x}
\end{align}
where ${\cal N}$ is some normalization constant and the effective action
$S_{\rm  eff}(k)$
is given by
\begin{align}
  S_{\rm  eff}(k)
  &=\frac{1}{2\gamma} k^2 -ik + \frac{1}{2} \log (ik+ \epsilon)
  + \frac{d}{2} \log (-ik+ \epsilon) \ .
  \label{effective-k}
\end{align}

In order to evaluate the integral \eqref{kxx-integrate-x}, let us use
the saddle-point method. The saddle-point equation
\begin{align}
  0 &= \frac{d S_{\rm  eff}(k)}{dk}
  =\frac{1}{\gamma} k -i + \frac{i}{2} \frac{1}{ik+ \epsilon}
  - \frac{id}{2} \frac{1}{-ik+ \epsilon}
\end{align}
has three solutions,
among which
there exists a solution
\begin{align}
  k^{(0)} \simeq i  \, \frac{d-1}{d+1} \epsilon  + i \frac{8d}{(d+1)^3} \epsilon^2
 + {\rm O}(\epsilon^3) \ ,
  \label{relevant-saddle-k}
\end{align}
which goes to zero as $\epsilon \rightarrow 0$.
Note that the denominators in \eqref{kxx-integrate-x} becomes
\begin{align}
  i k^{(0)} + \epsilon  \simeq
  \frac{2}{d+1} \epsilon  + {\rm O}(\epsilon^2) \ ,
  \\
  - i k^{(0)} + \epsilon  \simeq
  \frac{2d}{d+1} \epsilon  + {\rm O}(\epsilon^2) \ ,
\end{align}
at the saddle point.
Thus,
one finds that the partition function
diverges as
\begin{align}
  Z_{\epsilon} \sim \epsilon ^{-\frac{d+1}{2}}
  \label{diverging-partition-fn}
\end{align}
for $\epsilon \rightarrow 0$
if one ignores the fluctuations of $k$ around the saddle point.
(Strictly speaking, one gets an extra factor of ${\rm O}(\epsilon)$
from the fluctuations.)

Let us note here that, in the model \eqref{k-x-integral}, 
there is an identity such as
\begin{align}
  \langle k \rangle_{\epsilon}
  &=   i \gamma \langle (\eta_{\mu\nu }x_\mu x_\nu +1) \rangle_{\epsilon} \ ,
  \label{k-x2-identity}
\end{align}
which can be derived by changing the variable as
   $k\to k+i\gamma(\eta_{\mu\nu} x_\mu x_\nu+1)$
so that the linear term in $k$ in the action is eliminated.
On the other hand,
the left-hand side can be evaluated by
using the partition function \eqref{kxx-integrate-x}
obtained after integrating out $x$. 
Since the saddle point \eqref{relevant-saddle-k} makes the partition function
divergent as we have seen in \eqref{diverging-partition-fn},
the fluctuations around the saddle point are strongly suppressed
as $\epsilon \rightarrow 0$, which we prove in a separate paper \cite{largeD}.
Hence the left-hand side of \eqref{k-x2-identity} vanishes
in the $\epsilon \rightarrow 0$ limit, which implies
\begin{align}
  \lim_{\epsilon\rightarrow 0}
  \langle \eta_{\mu\nu} x_\mu x_\nu \rangle_{\epsilon} &= - 1 \ .
  \label{cut-off-model-classic}
\end{align}
This shows that there are no quantum corrections to this observable
even at finite $\gamma$.

\textit{The mechanism of classicalization.---}
In order to understand why the classicalization occurs
when we regularize the model \eqref{gaussian-quartic}
by the Lorentz symmetry breaking cutoff,
we discuss the fluctuations around the saddle points
on $\eta_{\mu\nu} x_\mu x_\nu = -1$,
which are related to each other by
Lorentz transformation.
For that, we consider the Hessian at each saddle point
\begin{align}
  H_{\mu \nu} &= \frac{\del ^2 S(x)}{\del x_\mu \del x_\nu}
  = \gamma \, \eta_{\mu\lambda} \eta_{\nu\rho} x_\lambda x_\rho \ ,
  \label{def-Hessian}
  \end{align}
  which is a real symmetric $(d+1)\times (d+1)$ matrix.

  Note first that,
  under the Lorentz transformation
\begin{align}
  x_\mu^{\cal O} &= {\cal O}_{\mu\nu }x_\nu  \ ,
  \label{def-O-transf}
\end{align}
where ${\cal O}\in {\rm SO}(d,1)$,
  the Hessian transforms as
\begin{align}
  H_{\mu\nu}(x)
  &= {\cal O}_{\lambda \mu} H_{\lambda \rho}(x^{\cal O})\,  {\cal O}_{\rho \nu} \ .
  \end{align}
The change of the action for the fluctuation $\delta x_\mu$
around the saddle point $x_\mu$ is
\begin{align}
  \delta S &=
\delta x_\mu  H_{\mu\nu}(x) \delta x_\nu
= \delta x_\mu '
H_{\mu \nu}(x^{\cal O})\,
\delta x_\nu '     \ ,
\label{delta-S-cov}
  \end{align}
where $\delta x_\mu ' = {\cal O}_{\mu \nu} \delta x_\nu$.
Thus the fluctuations get Lorentz boosted for the boosted saddle point.

\begin{figure}[t]
	\centering
	\includegraphics[width=0.45\textwidth]{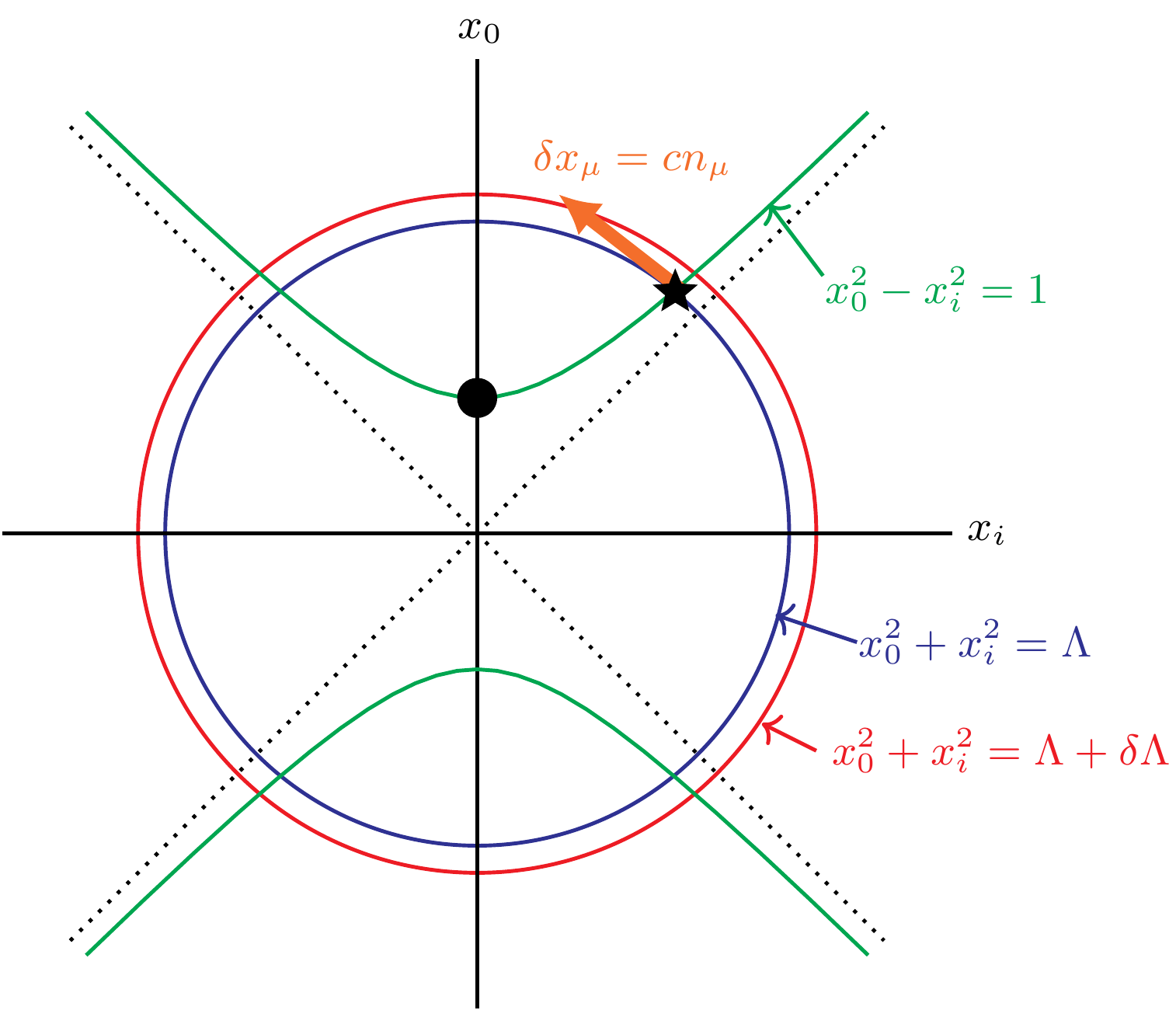}
	\caption{
          The fluctuations around the boosted saddle point (star),
     which can be obtained by applying a Lorentz
          transformation to the unboosted saddle point (circle),
          are restricted to the
          thin shell region between the cutoff surfaces, which leads to a steep increase of the
          action.}
        \label{fig:cutoff-fluct}
\end{figure}

Let us then consider what happens
when we introduce the Lorentz symmetry breaking cutoff \eqref{sym-br-cutoff}.
As $\epsilon$ gets smaller,
the integral is dominated by configurations around \emph{boosted} saddle points
as one can deduce from the fact that the partition function diverges.
However, the fluctuations around boosted saddle points are strongly affected by
the cutoff terms in \eqref{sym-br-cutoff}.
To simplify the argument, let us 
replace the cutoff in \eqref{sym-br-cutoff}
by a sharp cutoff given by $ (x_0)^2 +  (x_i)^2 \le \Lambda$.
See Fig.~\ref{fig:cutoff-fluct}.

As we increase the cutoff from $\Lambda$ to $\Lambda + \delta \Lambda$, 
we add a shell given by $ \Lambda \le (x_0)^2 +  (x_i)^2 \le \Lambda+ \delta \Lambda$
to the region of integration,
and we repeat this when we send $\Lambda$ to $\infty$.
Within each shell,
the fluctuations $\delta x_\mu$ around the saddle point
are restricted
to those satisfying  $x_0 \delta x_0  + x_i \delta x_i  = 0$.
Therefore the physical fluctuations around the
saddle point $x_\mu=(\cosh \sigma , \sinh \sigma , 0, \cdots , 0)$
are $\delta x_\mu= c \, n_\mu$,
where $c\in \bbR$
and we have defined a unit vector
$n =(\sinh \sigma , - \cosh \sigma , 0 , \cdots, 0 )/ \sqrt{\cosh(2\sigma)}$.

Plugging this in \eqref{delta-S-cov},
one obtains
the increase of the action
$\delta S = c^2 \gamma \sinh^2 (2 \sigma)/\cosh(2\sigma)$,
which becomes $c^2 \gamma e^{2\sigma}/2$ for large $\sigma$.
Therefore, the coefficient $c$ vanishes as
$|c|\lesssim e^{-\sigma}\sqrt{2/\gamma}$ for large $\sigma$,
which implies that
the fluctuations around the boosted saddle points are
strongly suppressed in the cutoff model.

\textit{``Gauge-fixing'' the Lorentz symmetry.---}
The previous discussions
suggest that
the Lorentz symmetry breaking cutoff affects the quantum fluctuations
around boosted saddle points drastically,
and such effects remain to be there even if one removes the cutoff.
On the other hand, the existence of Lorentz symmetry means that
all the points related to each other by Lorentz transformation
should be regarded as physically equivalent. If we consider
this symmetry as the guiding principle, the physically correct way
to define the integral is to reformulate the integral
so that
the equivalence class is represented by a unique point in each class.
Factoring out the divergent ``gauge volume'' of the Lorentz symmetry
in this way, 
we can make the partition function \eqref{gaussian-quartic} finite.
This can be achieved by
using the standard Faddeev-Popov gauge-fixing procedure.

 Let us note first that we can ``fix the gauge'' by minimizing $(x_0)^2$
with respect to
the Lorentz transformation ${\rm SO}(d,1)$.
In fact, there exists a unique minimum up to the ${\rm SO}(d)$ rotational symmetry,
which is characterized by the condition
\begin{align}
  x_0 x_i  &= 0  \quad \quad \mbox{for all~}i=1, \cdots , d \ .
  \label{gauge-fixing-cond}
\end{align}
Let us
use this as the gauge fixing condition
and
introduce the Faddeev-Popov (FP) determinant $\Delta_{\rm FP}$ by
\begin{align}
  \int d{\cal O} \, \Delta_{\rm FP}(x^{\cal O})
  \prod_{i=1}^d \delta(x_0^{\cal O} x_i^{\cal O}) &= 1 \ ,
  \label{FP-identity}
  \end{align}
where $x^{\cal O}$ is defined by \eqref{def-O-transf}.
Suppose ${\cal O}$ minimizes $(x_0^{\cal O})^2$ for a given $x_\mu$.
Then we consider how $x_0^{\cal O} x_i^{\cal O}$ changes under the Lorentz boost
\begin{align}
  x_0^{\cal O} (j,\sigma)&= x_0^{\cal O} \cosh \sigma + x_j^{\cal O} \sinh \sigma \ ,\\
  x_j^{\cal O} (j, \sigma) &= x_0^{\cal O} \sinh \sigma + x_j^{\cal O} \cosh \sigma \ ,\\
  x_k^{\cal O} (j, \sigma) &= x_k^{\cal O} \quad\quad   (\mbox{for~}k\neq j)
 \end{align}
in the $j$-direction. Let us define the $d\times d$ real symmetric matrix
\begin{align}
  \Omega_{ij}(x^{\cal O}) &= \left.
  \frac{d}{d\sigma} \left(
  x_0^{\cal O}(j,\sigma) \, x_i^{\cal O}(j,\sigma) \right) \right|_{\sigma=0} \\
  &= (x_0^{\cal O})^2 \delta_{ij} + x_i^{\cal O} x_j^{\cal O} \ .
\end{align}
Using this, 
the FP determinant
can be defined as 
\begin{align}
\Delta_{\rm FP}(x^{\cal O}) &= {\rm det} \, \Omega(x^{\cal O})  \ ,
\end{align}
if
the measure $d {\cal O}$ is defined with appropriate normalization.
Inserting the identity \eqref{FP-identity}
in the partition function \eqref{gaussian-quartic}
and using its Lorentz symmetry,
we obtain
\begin{align}
  Z_{\rm g.f.}  &= 
  \int \, dx  \, e^{-S(x)}  \Delta_{\rm FP}(x) \prod_{i=1}^d\delta(x_0 x_i)  \ ,
  \label{gauge-fixed-x-model}
\end{align}
where
the divergent ``gauge volume''
associated with the Lorentz symmetry is omitted.

In fact, the integral is dominated by $x_i=0$
and becomes
\begin{align}
  Z_{\rm g.f.} &=  \int \, dx_0  \, |x_0|^d \, e^{- \frac{1}{2} \gamma \{-(x_0)^2+1\}^2} \ .
\end{align}
Using this gauge-fixed partition function, we obtain
$\langle (x_0)^2 \rangle_{\rm g.f.} = 1 + (d-1)/(2 \gamma)+\cdots$.
This implies that the Lorentz invariant observable
\begin{align}
  \langle \eta_{\mu\nu} x_\mu x_\nu \rangle
  &= - 1 -
  \frac{d-1}{2\gamma} + \cdots
\label{xx-quantum}
\end{align}
has quantum corrections represented by the ${\rm O}(1/\gamma)$ terms,
which is
in striking contrast to the result \eqref{cut-off-model-classic}
obtained by the cutoff model. 
Namely
the cutoff model fails to reproduce the quantum corrections correctly.

\textit{Gaussian integral.---}
As an example in which the Lorentz symmetry breaking cutoff causes a more
subtle effect, let us consider the Gaussian integral
\begin{align}
  Z = \int \, dx \,  e^{\frac{1}{2} i \gamma \eta_{\mu\nu }x_\mu x_\nu}
   = \int \, dx \, e^{\frac{1}{2} i \gamma  \{- (x_0)^2 + (x_i)^2\} } \ ,
  \label{gaussian}
  \end{align}
where $\gamma>0$ and $x_\mu \in \bbR$ $(\mu = 0 , 1 , \cdots , d)$.
Since the integral is not absolutely convergent,
it is not well-defined as it is.
For instance, one can introduce a Lorentz symmetry breaking cutoff as
we did in \eqref{sym-br-cutoff}, and obtain
\begin{align}
  Z
  = e^{i(d-1) \pi/4}  \left( \frac{2 \pi}{\gamma} \right)^{(d+1)/2} 
  \label{result-gaussian}
\end{align}
in the $\epsilon \rightarrow 0$ limit.
Note that
we get a finite result in this case
despite the noncompact symmetry unlike in \eqref{gaussian-quartic}.
This is related to the fact that
the only saddle point of the integral \eqref{gaussian} is $x_\mu=0$,
which is invariant under the Lorentz transformation.
The classicalization due to the cutoff does not occur in this case. 

However,
if we restrict ourselves to configurations with 
$\eta_{\mu\nu} x_\mu x_{\nu}=C$ (constant) and
integrate over the Lorentz boost parameter first,
we clearly obtain divergence.
In order to define the model respecting the Lorentz symmetry,
let us ``gauge-fix'' the Lorentz symmetry as we did in \eqref{gauge-fixed-x-model}.
Then
the integral becomes
\begin{align}
  Z_{\rm g.f.} &=  \int \, dx_0  \, |x_0|^d   \, e^{- \frac{1}{2} i \gamma (x_0)^2 } \ .
  \label{result-gauge-fixed}
\end{align}
Since \eqref{result-gauge-fixed} is still not absolutely convergent,
we introduce the convergence factor $e^{-\epsilon (x_0)^2}$
as we did in \eqref{sym-br-cutoff} and take the $\epsilon \rightarrow 0$ limit
after integration.
This convergence factor is expected not to cause any problem
since the Lorentz symmetry is already gauge-fixed.
Thus one obtains $ Z_{\rm g.f.}\propto \gamma^{-(d+1)/2}$,
which is finite.
This conclusion clearly disagrees with the fact that the original integral
evaluated by the Lorentz symmetry breaking
regularization
is finite \eqref{result-gaussian}
since that would predict that the result after omitting the divergent ``gauge volume''
should be zero.
On the other hand,
the power of $\gamma$ obtained by the gauge-fixed model \eqref{result-gauge-fixed}
agrees with the cutoff model \eqref{result-gaussian}
simply on dimensional grounds,
and hence the Lorentz invariant observable turns out to be identical as
$\langle \eta_{\mu\nu} x_\mu x_\nu \rangle = i(d+1)/\gamma$.


\textit{Defining the type IIB matrix model.---}
The partition function of the type IIB matrix model can be written as
\begin{align}
  Z = \int d A \, e^{iS[A]} \, {\rm Pf}{\cal M}[A] \ ,
  \label{def-Z-ikkt}
\end{align}
where $A_\mu$ ($\mu = 0 , \cdots , 9$) are $N\times N$ traceless Hermitian
matrices and the action $S[A]$ is given by
\begin{align}
  S[A] &= - \frac{1}{4} N  \eta_{\mu\lambda} \eta_{\nu \rho} \,
  \tr [A_\mu , A_\nu][A_\lambda , A_\rho] \ .
\end{align}
The Pfaffian ${\rm Pf}{\cal M}[A] \in \bbR$
represents the contributions from the fermionic
matrices.
The model has ${\rm SO}(9,1)$ Lorentz symmetry $A_\mu ' = {\cal O}_{\mu\nu} A_\nu$
and the ${\rm SU}(N)$ symmetry $A_\mu ' = U A_\mu U^\dag$,
where $U\in {\rm SU}(N)$.

Since the integral \eqref{def-Z-ikkt} is not absolutely convergent,
one has to regularize it to make it well-defined.
In the old literature,
it was common to consider
the Euclidean version of the type IIB matrix model,
which can be obtained 
by deforming the integration contour as
$A_0 \mapsto e^{3 \pi i/8} A_0$ and $A_i \mapsto e^{-\pi i/8} A_i$.
The model one obtains in this way is ${\rm SO}(10)$ rotationally invariant
and totally well-defined \cite{9803117,0103159}.
Although intriguing spontaneous breaking of the SO(10) symmetry
to SO(3) occurs (See Ref.~\cite{Anagnostopoulos:2022dak} for a review.),
the emergent space-time is Euclidean.
We consider that
this is due to
the contour deformation used to define the model 
that breaks the Lorentz symmetry.

The Lorentzian version of the
type IIB matrix model
has been considered for the first time in Ref.~\cite{1108_1540}
using some Lorentz symmetry breaking cutoff.
After various trials and errors since then,
it was proposed to add a Lorentz-invariant mass
term \cite{Hatakeyama:2022ybs,Nishimura:2022alt} as
\begin{align}
S_{\gamma} &= 
-  \frac{1}{2} N \gamma \, \eta_{\mu\nu} \textrm{tr}  (A_{\mu} A_{\nu}) 
\end{align}
with $\gamma > 0$
and then to introduce convergence factors as
\begin{align}
  S_{\gamma}^{(\epsilon, \tilde{\epsilon})}
  &= \frac{1}{2} N \gamma \, \{ e^{i \tilde{\epsilon}}
  \textrm{tr} (A_0)^2 - e^{-i \epsilon} \textrm{tr} (A_i)^2 \} \ ,
  \label{gamma-epsilon-term}
\end{align}
which breaks Lorentz symmetry.
However, as our discussions above suggest,
this Lorentz symmetry breaking
may leave a severe artifact even if one takes
the $\epsilon, \tilde{\epsilon}\rightarrow 0$ limit later.
Indeed this will be demonstrated explicitly in the $N=2$ bosonic model
in the separate paper \cite{largeD}.


In order to perform numerical simulations of the model,
one applies either the complex Langevin method
(CLM) \cite{1904_05919,Hatakeyama:2022ybs,Nishimura:2022alt,Anagnostopoulos:2022dak} or the generalized
Lefschetz thimble method (GTM)
to overcome the sign problem that occurs due to the
complex integrand of the partition function.
However,
in these methods,
the SO(9,1) Lorentz symmetry is broken
by the noise term in the CLM and the flow equation in the GTM, respectively,
although the ${\rm SO}(9)(\subset {\rm SO}(9,1))$ symmetry is kept intact.
Therefore,
these methods fail to sample the boosted configurations
with the correct weight and the result
may be interpreted
effectively as that of the cutoff model with some
$\epsilon$ and $\tilde{\epsilon}$.

This motivates us to define
the type IIB matrix model by gauge-fixing the Lorentz symmetry
in a fully nonperturbative manner
as we did in the simple models.
Since the matrix configurations $A_\mu$ which are related to each other
by Lorentz transformation
should be regarded as physically equivalent, we pick up the unique
representative configuration (up to rotational symmetry) by minimizing
$\tr (A_0)^2$.
Generalizing the derivation of
\eqref{gauge-fixed-x-model} respecting the ${\rm SU}(N)$ invariance,
we
arrive at
\begin{align}
Z_{\rm g.f.} &=
  \int d A \,
  e^{iS[A]} \, {\rm Pf}{\cal M}(A)
\,   {\Delta}_{\rm FP}[A] \, \prod_{i=1}^d\delta(\tr (A_0 A_i))
  \ ,
  \label{gauge-fixed-ikkt}
\end{align}
where the FP determinant ${\Delta}_{\rm FP}[A]$
is given by
\begin{align}
  {\Delta}_{\rm FP}[A] &= {\rm det} \,  \Omega \ , \quad
  \Omega_{ij} = \tr (A_0)^2 \delta_{ij} + \tr (A_i A_j) \ .
\end{align}
Note that this model still
has ${\rm SO}(9)$ rotational symmetry $A_i \mapsto O_{ij} A_j$
($O\in {\rm SO}(9)$), under which
$\Omega \mapsto O \, \Omega \, O^\top$.
Using the eigenvalues $\lambda_i$ of $T_{ij}= \tr (A_i A_j)$,
we find ${\rm det}\, \Omega = \prod_{i=1}^{9} ( \tr (A_0)^2 + \lambda_i) \ge  0$.

Since the gauge-fixed model \eqref{gauge-fixed-ikkt}
is still not absolutely convergent,
we propose to introduce the convergence factors \eqref{gamma-epsilon-term}.
Note, however, that since the Lorentz symmetry is already gauge-fixed,
these
convergence factors are expected not to cause any problem.


\textit{Discussions.---}
What we discussed
in this Letter is, in fact,
quite general in that
it applies to any model that
has divergent partition function 
due to the existence of 
a noncompact symmetry.
Rather surprisingly, we find that
such a symmetry has to
be ``gauge-fixed’'
in order to define the model
without breaking the noncompact symmetry.
This is quite different from the situation in gauge theories
with a compact gauge group,
which can be defined on the lattice \emph{without fixing the gauge}.

Let us point out that
the dominant configurations
in the model with the gauge-fixed Lorentz symmetry
may well be very different from those
in the gauge-unfixed model (with some regularization)
since
the Faddeev-Popov determinant in \eqref{gauge-fixed-ikkt}
induces a new term in the saddle-point equation.
For instance, the commuting matrices satisfying $[A_\mu , A_\nu]=0$
are the saddle points in the original type IIB matrix model ($\gamma=0$),
but they are no longer saddle points in the model with
the gauge-fixed Lorentz symmetry.
It is therefore
important to perform
numerical simulations of the gauge-fixed
model \eqref{gauge-fixed-ikkt} proposed in this Letter
in order to
elucidate the nonperturbative dynamics of superstring theory
such as the emergence of (3+1)-dimensional
space-time.

When one simulates the
gauge-unfixed model,
one typically generates Lorentz boosted configurations \cite{Hirasawa:2024dht},
in which time and space are mixed up.
In contrast, when one simulates the gauge-fixed model,
the redundancy due to the Lorentz symmetry is taken into account completely
by generating only the ``unboosted configurations'' that
minimize $\tr (A_0)^2$,
which enables us to identify the eigenvalues of $A_0$ as the
``time coordinates''.

This is important in performing numerical simulations
in such a way that the time-evolution can be extracted
from the generated configurations \cite{1108_1540}.
For that, we use the ${\rm SU}(N)$ symmetry to make $A_0$
into a diagonal form $A_0 = {\rm diag} (\alpha_1 , \cdots , \alpha_N)$,
where $\alpha_1 < \cdots < \alpha _N$.
If $A_i$ $(i = 1 , \cdots , 9)$ have band-diagonal structure
with the band width $n$ in this basis, we can
define the $n\times n$ submatrices
$(\bar{A}_i)_{IJ}(t_a)\equiv (A_i)_{a+I, a+J}$ ($I,J=1,\cdots , n$),
which represent the nine-dimensional space at each time
$t_a = \sum_{b=a+1}^{a+n}\alpha_b$.
When we apply the CLM or GTM to simulate this model,
we have to complexify the dynamical variables $\alpha_a$ and $A_i$
separately \cite{1904_05919}.
The expectation values of $\alpha_a$
represent
the time coordinates
and
the emergence of (3+1)-dimensional space-time
can be investigated
by looking at $T_{ij}(t)\equiv {\rm Tr} \Big( \bar{A}_i(t) \bar{A}_j(t) \Big)$.
Note that this does not work if Lorentz boosts occur during the simulation
as in the gauge-unfixed model.

Last but not the least, 
the gauge-unfixed model is, in fact, not easy to simulate since it
requires very long time to sample boosted configurations with the correct weight,
which is completely avoided in the gauge-fixed model.
Implementing the gauge-fixing condition and the Faddeev-Popov determinant
in the simulation is straightforward and the extra computational cost is negligible.



\textit{Acknowledgments.---}
We thank Chien-Yu Chou, Hikaru Kawai, Harold Steinacker
and Ashutosh Tripathi for valuable discussions.
J.N.~is grateful to the coauthors of Ref.~\cite{Anagnostopoulos:2022dak}
for a long-term collaboration, which partly motivated this work.

\bibliography{def-ikktv15}

\begin{thebibliography}{13}%
\makeatletter
\providecommand \@ifxundefined [1]{%
 \@ifx{#1\undefined}
}%
\providecommand \@ifnum [1]{%
 \ifnum #1\expandafter \@firstoftwo
 \else \expandafter \@secondoftwo
 \fi
}%
\providecommand \@ifx [1]{%
 \ifx #1\expandafter \@firstoftwo
 \else \expandafter \@secondoftwo
 \fi
}%
\providecommand \natexlab [1]{#1}%
\providecommand \enquote  [1]{``#1''}%
\providecommand \bibnamefont  [1]{#1}%
\providecommand \bibfnamefont [1]{#1}%
\providecommand \citenamefont [1]{#1}%
\providecommand \href@noop [0]{\@secondoftwo}%
\providecommand \href [0]{\begingroup \@sanitize@url \@href}%
\providecommand \@href[1]{\@@startlink{#1}\@@href}%
\providecommand \@@href[1]{\endgroup#1\@@endlink}%
\providecommand \@sanitize@url [0]{\catcode `\\12\catcode `\$12\catcode `\&12\catcode `\#12\catcode `\^12\catcode `\_12\catcode `\%12\relax}%
\providecommand \@@startlink[1]{}%
\providecommand \@@endlink[0]{}%
\providecommand \url  [0]{\begingroup\@sanitize@url \@url }%
\providecommand \@url [1]{\endgroup\@href {#1}{\urlprefix }}%
\providecommand \urlprefix  [0]{URL }%
\providecommand \Eprint [0]{\href }%
\providecommand \doibase [0]{https://doi.org/}%
\providecommand \selectlanguage [0]{\@gobble}%
\providecommand \bibinfo  [0]{\@secondoftwo}%
\providecommand \bibfield  [0]{\@secondoftwo}%
\providecommand \translation [1]{[#1]}%
\providecommand \BibitemOpen [0]{}%
\providecommand \bibitemStop [0]{}%
\providecommand \bibitemNoStop [0]{.\EOS\space}%
\providecommand \EOS [0]{\spacefactor3000\relax}%
\providecommand \BibitemShut  [1]{\csname bibitem#1\endcsname}%
\let\auto@bib@innerbib\@empty
\bibitem [{\citenamefont {Ishibashi}\ \emph {et~al.}(1997)\citenamefont {Ishibashi}, \citenamefont {Kawai}, \citenamefont {Kitazawa},\ and\ \citenamefont {Tsuchiya}}]{9612115}%
  \BibitemOpen
  \bibfield  {author} {\bibinfo {author} {\bibfnamefont {N.}~\bibnamefont {Ishibashi}}, \bibinfo {author} {\bibfnamefont {H.}~\bibnamefont {Kawai}}, \bibinfo {author} {\bibfnamefont {Y.}~\bibnamefont {Kitazawa}},\ and\ \bibinfo {author} {\bibfnamefont {A.}~\bibnamefont {Tsuchiya}},\ }\bibfield  {title} {\bibinfo {title} {{A Large N reduced model as superstring}},\ }\href {https://doi.org/10.1016/S0550-3213(97)00290-3} {\bibfield  {journal} {\bibinfo  {journal} {Nucl. Phys. B}\ }\textbf {\bibinfo {volume} {498}},\ \bibinfo {pages} {467} (\bibinfo {year} {1997})},\ \Eprint {https://arxiv.org/abs/hep-th/9612115} {arXiv:hep-th/9612115} \BibitemShut {NoStop}%
\bibitem [{\citenamefont {Anagnostopoulos}\ \emph {et~al.}(2023)\citenamefont {Anagnostopoulos}, \citenamefont {Azuma}, \citenamefont {Hatakeyama}, \citenamefont {Hirasawa}, \citenamefont {Ito}, \citenamefont {Nishimura}, \citenamefont {Papadoudis},\ and\ \citenamefont {Tsuchiya}}]{Anagnostopoulos:2022dak}%
  \BibitemOpen
  \bibfield  {author} {\bibinfo {author} {\bibfnamefont {K.~N.}\ \bibnamefont {Anagnostopoulos}}, \bibinfo {author} {\bibfnamefont {T.}~\bibnamefont {Azuma}}, \bibinfo {author} {\bibfnamefont {K.}~\bibnamefont {Hatakeyama}}, \bibinfo {author} {\bibfnamefont {M.}~\bibnamefont {Hirasawa}}, \bibinfo {author} {\bibfnamefont {Y.}~\bibnamefont {Ito}}, \bibinfo {author} {\bibfnamefont {J.}~\bibnamefont {Nishimura}}, \bibinfo {author} {\bibfnamefont {S.~K.}\ \bibnamefont {Papadoudis}},\ and\ \bibinfo {author} {\bibfnamefont {A.}~\bibnamefont {Tsuchiya}},\ }\bibfield  {title} {\bibinfo {title} {{Progress in the numerical studies of the type IIB matrix model}},\ }\href {https://doi.org/10.1140/epjs/s11734-023-00849-x} {\bibfield  {journal} {\bibinfo  {journal} {Eur. Phys. J. ST}\ }\textbf {\bibinfo {volume} {232}},\ \bibinfo {pages} {3681} (\bibinfo {year} {2023})},\ \Eprint {https://arxiv.org/abs/2210.17537} {arXiv:2210.17537 [hep-th]} \BibitemShut {NoStop}%
\bibitem [{\citenamefont {Brahma}\ \emph {et~al.}(2022)\citenamefont {Brahma}, \citenamefont {Brandenberger},\ and\ \citenamefont {Laliberte}}]{Brahma:2022ikl}%
  \BibitemOpen
  \bibfield  {author} {\bibinfo {author} {\bibfnamefont {S.}~\bibnamefont {Brahma}}, \bibinfo {author} {\bibfnamefont {R.}~\bibnamefont {Brandenberger}},\ and\ \bibinfo {author} {\bibfnamefont {S.}~\bibnamefont {Laliberte}},\ }\bibfield  {title} {\bibinfo {title} {{BFSS Matrix Model Cosmology: Progress and Challenges}},\ }\href@noop {} {\  (\bibinfo {year} {2022})},\ \Eprint {https://arxiv.org/abs/2210.07288} {arXiv:2210.07288 [hep-th]} \BibitemShut {NoStop}%
\bibitem [{\citenamefont {Klinkhamer}(2023)}]{Klinkhamer:2022frp}%
  \BibitemOpen
  \bibfield  {author} {\bibinfo {author} {\bibfnamefont {F.~R.}\ \bibnamefont {Klinkhamer}},\ }\bibfield  {title} {\bibinfo {title} {{Emergent gravity from the IIB matrix model and cancellation of a cosmological constant}},\ }\href {https://doi.org/10.1088/1361-6382/accef5} {\bibfield  {journal} {\bibinfo  {journal} {Class. Quant. Grav.}\ }\textbf {\bibinfo {volume} {40}},\ \bibinfo {pages} {124001} (\bibinfo {year} {2023})},\ \Eprint {https://arxiv.org/abs/2212.00709} {arXiv:2212.00709 [hep-th]} \BibitemShut {NoStop}%
\bibitem [{\citenamefont {Steinacker}(2024)}]{Steinacker:2024unq}%
  \BibitemOpen
  \bibfield  {author} {\bibinfo {author} {\bibfnamefont {H.~C.}\ \bibnamefont {Steinacker}},\ }\href {https://doi.org/10.1017/9781009440776} {\emph {\bibinfo {title} {{Quantum Geometry, Matrix Theory, and Gravity}}}}\ (\bibinfo  {publisher} {Cambridge University Press},\ \bibinfo {year} {2024})\BibitemShut {NoStop}%
\bibitem [{\citenamefont {Asano}\ \emph {et~al.}()\citenamefont {Asano}, \citenamefont {Nishimura}, \citenamefont {Piensuk},\ and\ \citenamefont {Yamamori}}]{largeD}%
  \BibitemOpen
  \bibfield  {author} {\bibinfo {author} {\bibfnamefont {Y.}~\bibnamefont {Asano}}, \bibinfo {author} {\bibfnamefont {J.}~\bibnamefont {Nishimura}}, \bibinfo {author} {\bibfnamefont {W.}~\bibnamefont {Piensuk}},\ and\ \bibinfo {author} {\bibfnamefont {N.}~\bibnamefont {Yamamori}},\ }\bibinfo {title} {{in preparation}}\BibitemShut {NoStop}%
\bibitem [{\citenamefont {Krauth}\ \emph {et~al.}(1998)\citenamefont {Krauth}, \citenamefont {Nicolai},\ and\ \citenamefont {Staudacher}}]{9803117}%
  \BibitemOpen
\bibfield  {title} {  }\bibfield  {author} {\bibinfo {author} {\bibfnamefont {W.}~\bibnamefont {Krauth}}, \bibinfo {author} {\bibfnamefont {H.}~\bibnamefont {Nicolai}},\ and\ \bibinfo {author} {\bibfnamefont {M.}~\bibnamefont {Staudacher}},\ }\bibfield  {title} {\bibinfo {title} {{Monte Carlo approach to M theory}},\ }\href {https://doi.org/10.1016/S0370-2693(98)00557-7} {\bibfield  {journal} {\bibinfo  {journal} {Phys. Lett. B}\ }\textbf {\bibinfo {volume} {431}},\ \bibinfo {pages} {31} (\bibinfo {year} {1998})},\ \Eprint {https://arxiv.org/abs/hep-th/9803117} {arXiv:hep-th/9803117} \BibitemShut {NoStop}%
\bibitem [{\citenamefont {Austing}\ and\ \citenamefont {Wheater}(2001)}]{0103159}%
  \BibitemOpen
  \bibfield  {author} {\bibinfo {author} {\bibfnamefont {P.}~\bibnamefont {Austing}}\ and\ \bibinfo {author} {\bibfnamefont {J.~F.}\ \bibnamefont {Wheater}},\ }\bibfield  {title} {\bibinfo {title} {{Convergent Yang-Mills matrix theories}},\ }\href {https://doi.org/10.1088/1126-6708/2001/04/019} {\bibfield  {journal} {\bibinfo  {journal} {JHEP}\ }\textbf {\bibinfo {volume} {04}},\ \bibinfo {pages} {019}},\ \Eprint {https://arxiv.org/abs/hep-th/0103159} {arXiv:hep-th/0103159} \BibitemShut {NoStop}%
\bibitem [{\citenamefont {Kim}\ \emph {et~al.}(2012)\citenamefont {Kim}, \citenamefont {Nishimura},\ and\ \citenamefont {Tsuchiya}}]{1108_1540}%
  \BibitemOpen
  \bibfield  {author} {\bibinfo {author} {\bibfnamefont {S.-W.}\ \bibnamefont {Kim}}, \bibinfo {author} {\bibfnamefont {J.}~\bibnamefont {Nishimura}},\ and\ \bibinfo {author} {\bibfnamefont {A.}~\bibnamefont {Tsuchiya}},\ }\bibfield  {title} {\bibinfo {title} {{Expanding (3+1)-dimensional universe from a Lorentzian matrix model for superstring theory in (9+1)-dimensions}},\ }\href {https://doi.org/10.1103/PhysRevLett.108.011601} {\bibfield  {journal} {\bibinfo  {journal} {Phys. Rev. Lett.}\ }\textbf {\bibinfo {volume} {108}},\ \bibinfo {pages} {011601} (\bibinfo {year} {2012})},\ \Eprint {https://arxiv.org/abs/1108.1540} {arXiv:1108.1540 [hep-th]} \BibitemShut {NoStop}%
\bibitem [{\citenamefont {Hatakeyama}\ \emph {et~al.}(2022)\citenamefont {Hatakeyama}, \citenamefont {Anagnostopoulos}, \citenamefont {Azuma}, \citenamefont {Hirasawa}, \citenamefont {Ito}, \citenamefont {Nishimura}, \citenamefont {Papadoudis},\ and\ \citenamefont {Tsuchiya}}]{Hatakeyama:2022ybs}%
  \BibitemOpen
  \bibfield  {author} {\bibinfo {author} {\bibfnamefont {K.}~\bibnamefont {Hatakeyama}}, \bibinfo {author} {\bibfnamefont {K.}~\bibnamefont {Anagnostopoulos}}, \bibinfo {author} {\bibfnamefont {T.}~\bibnamefont {Azuma}}, \bibinfo {author} {\bibfnamefont {M.}~\bibnamefont {Hirasawa}}, \bibinfo {author} {\bibfnamefont {Y.}~\bibnamefont {Ito}}, \bibinfo {author} {\bibfnamefont {J.}~\bibnamefont {Nishimura}}, \bibinfo {author} {\bibfnamefont {S.}~\bibnamefont {Papadoudis}},\ and\ \bibinfo {author} {\bibfnamefont {A.}~\bibnamefont {Tsuchiya}},\ }\bibfield  {title} {\bibinfo {title} {{Complex Langevin studies of the emergent space-time in the type IIB matrix model}}\ }(\bibinfo {year} {2022})\ \Eprint {https://arxiv.org/abs/2201.13200} {arXiv:2201.13200 [hep-th]} \BibitemShut {NoStop}%
\bibitem [{\citenamefont {Nishimura}(2022)}]{Nishimura:2022alt}%
  \BibitemOpen
  \bibfield  {author} {\bibinfo {author} {\bibfnamefont {J.}~\bibnamefont {Nishimura}},\ }\bibfield  {title} {\bibinfo {title} {{Signature change of the emergent space-time in the IKKT matrix model}},\ }\href {https://doi.org/10.22323/1.406.0255} {\bibfield  {journal} {\bibinfo  {journal} {PoS}\ }\textbf {\bibinfo {volume} {CORFU2021}},\ \bibinfo {pages} {255} (\bibinfo {year} {2022})},\ \Eprint {https://arxiv.org/abs/2205.04726} {arXiv:2205.04726 [hep-th]} \BibitemShut {NoStop}%
\bibitem [{\citenamefont {Nishimura}\ and\ \citenamefont {Tsuchiya}(2019)}]{1904_05919}%
  \BibitemOpen
  \bibfield  {author} {\bibinfo {author} {\bibfnamefont {J.}~\bibnamefont {Nishimura}}\ and\ \bibinfo {author} {\bibfnamefont {A.}~\bibnamefont {Tsuchiya}},\ }\bibfield  {title} {\bibinfo {title} {{Complex Langevin analysis of the space-time structure in the Lorentzian type IIB matrix model}},\ }\href {https://doi.org/10.1007/JHEP06(2019)077} {\bibfield  {journal} {\bibinfo  {journal} {JHEP}\ }\textbf {\bibinfo {volume} {06}},\ \bibinfo {pages} {077}},\ \Eprint {https://arxiv.org/abs/1904.05919} {arXiv:1904.05919 [hep-th]} \BibitemShut {NoStop}%
\bibitem [{\citenamefont {Hirasawa}\ \emph {et~al.}(2024)\citenamefont {Hirasawa}, \citenamefont {Anagnostopoulos}, \citenamefont {Azuma}, \citenamefont {Hatakeyama}, \citenamefont {Nishimura}, \citenamefont {Papadoudis},\ and\ \citenamefont {Tsuchiya}}]{Hirasawa:2024dht}%
  \BibitemOpen
  \bibfield  {author} {\bibinfo {author} {\bibfnamefont {M.}~\bibnamefont {Hirasawa}}, \bibinfo {author} {\bibfnamefont {K.~N.}\ \bibnamefont {Anagnostopoulos}}, \bibinfo {author} {\bibfnamefont {T.}~\bibnamefont {Azuma}}, \bibinfo {author} {\bibfnamefont {K.}~\bibnamefont {Hatakeyama}}, \bibinfo {author} {\bibfnamefont {J.}~\bibnamefont {Nishimura}}, \bibinfo {author} {\bibfnamefont {S.}~\bibnamefont {Papadoudis}},\ and\ \bibinfo {author} {\bibfnamefont {A.}~\bibnamefont {Tsuchiya}},\ }\bibfield  {title} {\bibinfo {title} {{The effects of SUSY on the emergent spacetime in the Lorentzian type IIB matrix model}},\ }in\ \href@noop {} {\emph {\bibinfo {booktitle} {{23rd Hellenic School and Workshops on Elementary Particle Physics and Gravity}}}}\ (\bibinfo {year} {2024})\ \Eprint {https://arxiv.org/abs/2407.03491} {arXiv:2407.03491 [hep-th]} \BibitemShut {NoStop}%
\end{thebibliography}%


\end{document}